\documentclass[prl,twocolumn,showpacs,aps]{revtex4}
\usepackage{graphics}
\usepackage[sort&compress]{natbib}
\newcommand{\bb}{{\mathbf{b}}}
\newcommand{\rb}{{\mathbf{r}}}
\newcommand{\Rb}{{\mathbf{R}}}
\newcommand{\Ec}{{\cal E}}
\newcommand{\mPhi}{{\mit\Phi}}

\newcommand{\sump}{\mathop{{\sum}'}}
\begin{document}
\title{Local lattice potentials and steady-state vacancies in ionic
crystals}
\author{Eugene V. Kholopov}
\email{kholopov@casper.che.nsk.su}
\affiliation{Institute of Inorganic Chemistry of the Siberian Branch of
the Russian Academy of Sciences, 630090 Novosibirsk, Russia}
\begin{abstract}
Basing on peculiarities of local potentials, the two principal trends in
vacancy formation are revealed. The proximity to the threshold of local
ionic stability due to a giant potential contribution of electronic
delocalization accounts for thermal anion vacancies typical of intrinsic
semiconductors of $AB$ type. On the other hand, the tendency towards
equalizing the potential field results in a high concentration of
structural cation vacancies observed in Ni$_3$Sb and relative compounds.
\end{abstract}
\maketitle

Bulk electrostatic potential field plays a fundamental role in crystals
\cite{Spac81}, for it specifies both the electro\-negativity
\cite{Spac81,Boyd81,Gene87} and charge transfer \cite{Wang80,Sato81,%
Cole97,Koha98} and eventually contributes to the cohesive energy there
\cite{Wadd59,Tosi64,Tref87}. Described by lattice series \cite{Wadd59,%
Tosi64,Harr75,Glas80}, bulk potentials can nevertheless be defined as
unique quantities subjected to invariant periodic boundary conditions
\cite{Khol01,Kho02a}. The particular event of point charge lattices is of
special interest \cite{Made18,Evje32,Bert52,HarMon,Bert78,Herz79,%
Heye81,Mass82,Rein90,Goni91,Argy92,Vito95} because it is relevant to
spherically symmetric ions of which external potential effect amounts to
that of point charges \cite{Jack62}. However, the symmetry of ionic
potentials evident in this case \cite{Spac81,Roy954,Birm55,%
Levi66,Goo69a,Cala76,Xiao92,Kho02b} cannot explain the known fact that
the steady-state concentrations of vacancies of various ionic species
differ even in the simplest case of diatomic crystals \cite{Goo69b}. The
nature of abnormal concentrations of cation vacancies in Ni$_3$Sb and
relative compounds \cite{Robe91,Rand96,Best98} is one more intriguing
problem.

In the present paper we discuss in detail that the effect of prevailing
anion vacancies inherent in GaAs \cite{Bara86,Nort94,Oate95,Jano97} as
well as in structurally relative nitrides \cite{Bogu95,Piqu97,Gorc99,%
Orel01} can be understood if the potential contribution of electronic
delocalization \cite{Khol01,Kho02b,Kho02c} is taken into account. But it
turns out that another treatment based on lattice potentials is
appropriate to explain cation vacancies in question.

To discuss local properties in ionic crystals, we consider a crystal
composed of $j$ ionic species per unit cell, with basis vectors $\bb_j$
and with total charges $q_j$ described by charge distributions
$\rho_j(\rb)$ spherically symmetric for simplicity. According to
\cite{Khol01,Kho02a,Kho02b}, the bulk electrostatic potential $U_{\mathrm
b}(\rb)$ at a point $\rb$ can be written as follows:
\begin{equation}\label{Aq1}
U_{\mathrm b}(\rb)=U_{\mathrm{Cd}}(\rb)+\mPhi_{\mathrm{top}} ,
\end{equation}
where
\begin{eqnarray}
&{\displaystyle U_{\mathrm{Cd}}(\rb)=\sump_i\sum_j\int_{V_j}\frac{\rho_j
(\rb')\:d\rb'}{|\rb'+\Rb_i+\bb_j-\rb|}} ,&\label{Aq2}\\
&{\displaystyle\mPhi_{\mathrm{top}}=\frac{2\pi}{3v}\sum_j\int_{V_j}|\rb
+\bb_j|^2\rho_j(\rb)\:d\rb} ,&\label{Aq3}
\end{eqnarray}
the sum over $i$ runs over the Bravais lattice specified by $\Rb_i$, the
prime on the summation sign implies missing the singularities of the
summand, $V_j$ is the volume of integration appropriate to $\rho_j(\rb)$,
$v$ is the volume of the unit cell. It is important that the absolute
convergence of interest can always be achieved for (\ref{Aq2}) by
introducing fictitious point charges, which in turn can be included into
the set over $j$ \cite{Khol01,Kho02a}. Note that if the distributions
$\rho_j(\rb)$ are non-overlapping, then (\ref{Aq2}) describes the
potential in the point-charge lattice, but (\ref{Aq3}) can be represented
as
\begin{eqnarray}
&{\displaystyle\mPhi_{\mathrm{top}}=\frac{2\pi}{3v}\sum_j|\bb_j|^2q_j
+\mPhi_{\mathrm{top}}^{\mathrm{deloc}}} ,&\label{Aq4}\\
&{\displaystyle\mPhi_{\mathrm{top}}^{\mathrm{deloc}}=\frac{2\pi}{3v}\sum_j
\int_{V_j}|\rb|^2\rho_j(\rb)\,d\rb} ,&\label{Aq5}
\end{eqnarray}
where the first term on the right-hand side of (\ref{Aq4}) modifies the
result of (\ref{Aq2}) so as to retrieve the potential symmetry in a
point-charge lattice, but $\mPhi_{\mathrm{top}}^{\mathrm{deloc}}$
determines the potential contribution of electronic delocalization in
ions, providing that this potential shift is intrinsically negative and
additive with respect to all ions in question \cite{Khol01,Kho02b}.

Based on (\ref{Aq1}), the bulk Coulomb energy takes the form
\begin{equation}\label{Aq6}
\Ec=\frac{1}{2}\sum_j q_j\,U_j^{\mathrm{eff}} ,
\end{equation}
where the contribution of fictitious charges mentioned above vanishes by
definition, so that $j$ runs over all the actual ions in the unit cell
\cite{Khol01,Kho02b},
\begin{equation}\label{Aq7}
U_j^{\mathrm{eff}}=\frac{1}{q_j}\int_{V_j}\rho_j(\rb)\:U_{\mathrm b}
(\rb+\bb_j)\,d\rb .
\end{equation}
In the foregoing case of point-like charges, $U_j^{\mathrm{eff}}$ are
reduced to the corresponding point-charge-lattice ionic potentials $U^0_j$
\cite{Khol01,Kho02b}, but shifted by
$\mPhi_{\mathrm{top}}^{\mathrm{deloc}}$.

It is significant that every ion should be bound energetically in the
lattice. This claim implies that
\begin{equation}\label{Aq8}
q_j\,U_j^{\mathrm{eff}}<0 ,
\end{equation}
despite the fact that according to (\ref{Aq6}) and the overall neutrality
of the unit cell, the entire value of $\Ec_{\mathrm b}$ is indifferent to
any uniform shift of the potential values. Hence, negative ions are to
stay in positive potential fields, so that the admissible negative value
of $\mPhi_{\mathrm{top}}^{\mathrm{deloc}}$ appears to be restricted for
every actual ionic structure and the lattice instability should be
expected otherwise.

Note that $\mPhi_{\mathrm top}^{\mathrm deloc}$ does not change relative
values on the bulk potential map, but describes the binding energies
relative to free ions, with including their own stability
\cite{Wats58,Doll95,Stef98}. Therefore, it can be essential in the
problem of steady-state vacancies forming at high temperatures. To
estimate this effect, let a vacancy be generated by the simple exclusion
of an ion from its regular position without any relaxation of the
surroundings \cite{HarMon,Goo69b,Lesl85,Bhow88,Ston88}. The total
concentration of vacancies is assumed to be small enough, so that their
influence on the bulk potentials \cite{Cole97,Macd82,Wolv95} can be
ignored at least while we are interested in thermal vacancies. The energy
loss appropriate to a vacancy of the $j$th species is then as follows:
\begin{equation}\label{Aq9}
E_j^{\mathrm{vac}}=-q_j\,U_j^{\mathrm{eff}} .
\end{equation}
Based on (\ref{Aq9}), the steady-state concentrations $n_j$ of vacant
states can be defined by the Boltzmann relation as
\begin{equation}\label{Aq10}
n_j=w_j\exp\Bigl(-\frac{E_j^{\mathrm{vac}}}{k_{\mathrm B}T}\Bigr) ,
\end{equation}
where $w_j$ is the number of energetically equivalent ions of the $j$th
species in the unit cell, $k_{\mathrm{B}}$ is the Boltzmann constant, $T$
is the temperature. The inequality $E_j^{\mathrm{vac}}>0$ follows from
(\ref{Aq8})--(\ref{Aq10}) as a necessary condition for $n_j\ll1$.

To make the further discussion complete, we start from the simplest case
of $AB$ compounds described by point-charge lattices \cite{Kho02c}. Then
there are only two charge species with $q_+=-q_-=\kappa$, $w_\pm=1$, and
\begin{equation}\label{Aq11}
E_\pm^{\mathrm{vac}}=\kappa\bigl(U^0_-\mp
\mPhi_{\mathrm{top}}^{\mathrm{deloc}}\bigr) ,
\end{equation}
where $U^0_-$ is the point-charge-lattice potential on $q_-$. According
to (\ref{Aq10}) and (\ref{Aq11}), for the ratio $n_-/n_+$ we derive
\begin{equation}\label{Aq12}
\frac{n_-}{n_+}=\exp\Bigl(-\frac{2\kappa\,
\mPhi_{\mathrm{top}}^{\mathrm{deloc}}}{k_BT}\Bigr)>1 .
\end{equation}

Typical energies associated with (\ref{Aq10}) are of about 10$^5\,${K}, so
that all the values of $n_j$ are usually negligible. However, the
contribution of $\mPhi_{\mathrm{top}}^{\mathrm{deloc}}$ can shift a
minimal positive potential $U_-^{\mathrm{eff}}$ to threshold (\ref{Aq8})
and thus the value of $n_-$ becomes appreciable. Presumably, this is the
case in semiconductors such as GaAs. To evaluate this particular event, we
adopt the lattice spacing $a=0.56537\,$nm \cite{Wyck63} and $\kappa=3e$,
where $e$ is the charge of proton. On utilizing the potential values for
the ZnS structure from \cite{Kho02b}, the expected magnitude of
$n_-\approx10^{-3}$ at $T\approx1000\,${K} \cite{Oate95} appears if we
invoke $\mPhi_{\mathrm{top}}^{\mathrm{deloc}}=-3.757\,\kappa/a$.

\begin{figure}[t]
\resizebox{0.68\hsize}{!}{\includegraphics{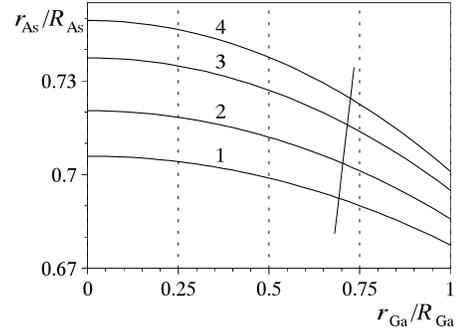}}
\caption{The ratio $r_{\mathrm{As}}/R_{\mathrm{As}}$ of a core ionic
radius to a crystal one for the As$^{3-}$ ion versus that for the
Ga$^{3+}$ ion so as to result in the value of
$\mPhi_{\mathrm{top}}^{\mathrm{deloc}}=-3.757\,\kappa/a$ expected for
GaAs. Curves 1, 2, 3, and 4 correspond to
$R_{\mathrm{Ga}}/R_{\mathrm{As}}=0.225$, 0.25, 0.279, and 0.3,
respectively. The thin solid line exhibits the case of
$r_{\mathrm{As}}/R_{\mathrm{As}}=r_{\mathrm{Ga}}/R_{\mathrm{Ga}}$.}
\label{Fig1}
\end{figure}

To estimate this value in terms of structural ionic parameters, we may
assume that there is a close-packed sphalerite structure composed of
spherical ions of radii $R_{\mathrm{Ga}}$ and $R_{\mathrm{As}}$, but their
core regions \cite{Birm55,Call58,Simm61,Olsz91} of radii $r_{\mathrm{Ga}}$
and $r_{\mathrm{As}}$, charged uniformly \cite{Bert78,Herz79,Argy92}
contribute to $\mPhi_{\mathrm{top}}^{\mathrm{deloc}}$. According to
(\ref{Aq5}), it is easy to get \cite{Khol01,Kho02b}
\begin{equation}\label{Aq13}
\mPhi_{\mathrm{top}}^{\mathrm{deloc}}=\frac{8\pi}{5a^3}\Bigl(
Z_{\mathrm{Ga}}r_{\mathrm{Ga}}^2+Z_{\mathrm{As}}r_{\mathrm{As}}^2\Bigr) ,
\end{equation}
where $Z_{\mathrm{Ga}}=-28e$ and $Z_{\mathrm{As}}=-36e$ are the total
electronic charges on ions. Keeping in mind the semiempirical character of
relationship (\ref{Aq13}), some variations in relative values of the ionic
radii are to be suggested upon comparing (\ref{Aq13}) with the value
$\mPhi_{\mathrm{top}}^{\mathrm{deloc}}$ of interest. The result is shown
in Fig.~\ref{Fig1}, where curve 1 corresponds to the limiting case when
the
As$^{3-}$ anions form a close-packed fcc structure themselves, curve 3 is
appropriate to the ratio $R_{\mathrm{Ga}}/R_{\mathrm{As}}=0.279$ proposed
by Pauling \cite{Paul60}.

It is instructive that a common relation
$r_{\mathrm{ion}}/R_{\mathrm{ion}}\approx0.7$ can be recognized in all the
cases depicted by the thin solid line in Fig.~\ref{Fig1}. According to
(\ref{Aq5}), in the particular event of
$\rho_{\mathrm{ion}}(\rb)=\exp(-|\rb|^2)$ appropriate to the powerful
\begin{figure}[b]
\resizebox{0.68\hsize}{!}{\includegraphics{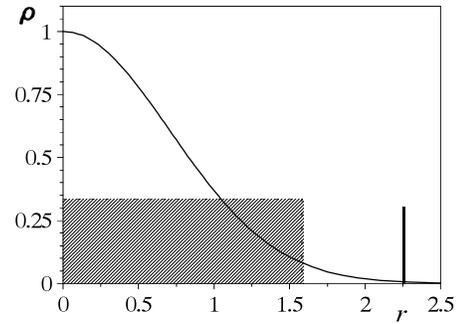}}
\caption{Gaussian charge distribution $\rho(r)=\exp(-r^2)$
(solid curve) and an equivalent uniform distribution (the hatched area)
restricted by $r=r_{\mathrm{ion}}$, whereas the vertical heavy line
corresponds to the crystal radius
$R_{\mathrm{ion}}=r_{\mathrm{ion}}/0.7$.}\label{Fig2}
\end{figure}
modern description of electronic configurations in atoms and ions in terms
of Gaussian orbitals \cite{Chel86,Cast98} we have
\begin{equation}\label{Aq14}
Z_{\mathrm{ion}}=\pi^{3/2} ,\qquad \mPhi_{\mathrm{top}}^{\mathrm{deloc}}
=\frac{\pi^{5/2}}{v} .
\end{equation}
The foregoing ionic parameters specified by (\ref{Aq13}) and (\ref{Aq14})
and imposed on this charge distribution in Fig.~\ref{Fig2} for comparison
are in favor of the above assessment.

Note that here we discuss the tendency towards arising vacancies in a
perfect system, but not their final states \cite{Bara86,Nort94,Oate95}.
Both the classical problem \cite{Goo69b,Koha00,Zhan01} of the $n$-type
conductivity in ZnO and that of vacancies in nitrides \cite{Bogu95,%
Piqu97,Gorc99} can be comprehended in the same fashion. The trends in
complex defects associated with anion vacancies \cite{Bara86,Jano97,%
Ling01} are clarified as well.

Owing to (\ref{Aq8}), the above mechanism enforces the stability of cation
states and so prevents from cation vacancies as a thermal effect. It means
that another nature brings about cation vacancies observed in a large
concentration in Ni$_3$Sb \cite{Rand96}. Of course, the experimental
existence of such a state implies the possibility of its thermodynamic
description with a large number of relevant parameters \cite{Best98}, but
the underlying physical motive remains obscure within such an approach.
To elucidate the reason promoting the appearance of this unusual state, we
resort to the potential effect associated with structural features.

\begin{figure}[t]
\resizebox{0.68\hsize}{!}{\includegraphics{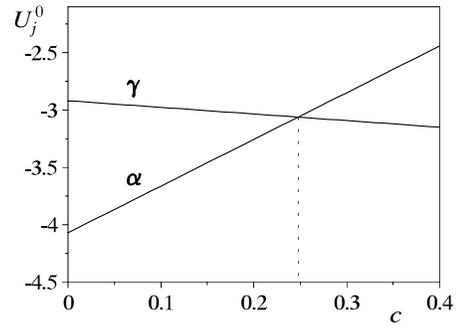}}
\caption{Local point-charge-lattice potentials $U^0_{\mathrm{\alpha}}$
(line $\alpha$) and $U^0_{\mathrm{\gamma}}$ (line $\gamma$) measured in
units of $\kappa/a$ versus the deviation $c$ from the total occupation of
$\alpha$ sites. The degenerate case pointed out by the dotted vertical
line at $c=0.248$ corresponds to the Ni$_{71.5}$Sb$_{28.5}$ composition.}
\label{Fig3}
\end{figure}

The stoichiometric Ni$_3$Sb compound has the D0$_3$ structure
\cite{Robe91,Rand96,Best98,Redd01} that is just the BiF$_3$ one
\cite{Wyck63}, with the local point-charge-lattice potentials listed in
Table~\ref{Table1} \cite{Khol01,Kho02b}, where the lattice parameter
$a\approx0.59\,$nm \cite{Rand96} and the charge $\kappa$ on the Ni$^+$ ion
\begin{table}
\caption{Local point-charge-lattice potentials, in units of $\kappa/a$,
describing Ni$_3$Sb within the D0$_3$ structure and the corresponding B1
potentials. Sb$^{3-}$ ions are on $\beta$ sites in the traditional
nomenclature \protect{\cite{Rand96}}, but Ni$^{+}$ ions occupy $\alpha$
and $\gamma$ sites specified in units of $a$ in the parentheses.}
\label{Table1}
\begin{ruledtabular}
\begin{tabular}{lrrr}
Structure&$U_{\beta}^0\:(0,\!0,\!0)$&$U_{\alpha}^0\:
(\frac{1}{4},\!\frac{1}{4},\!\frac{1}{4})\!$&$U_{\gamma}^0\:
(\frac{1}{2},\!0,\!0)$\vspace{0.5mm}\\
\hline
&&&\vspace{-3mm}\\
D0$_3\;$(BiF$_3$)&$11.06098140$&$-4.07072302$&$-2.91953536$\\
B1$\;$(NaCl)&$3.49512919$&$0\qquad$&$-3.49512919$
\end{tabular}
\end{ruledtabular}
\end{table}
are used as basic units. So, the $\alpha$ states of Ni$^+$ ions are
formally much more stable than the $\gamma$ ones. Notwithstanding, we must
propose that there are structural vacancies on the $\alpha$ sites, in
accord with experiment. The change in the local potentials can then be
written in the form
\begin{equation}\label{Aq15}
U^0_j=(1-c)U^0_j(\mbox{D}0_3)+c\,U^0_j(\mbox{B1}) ,
\end{equation}
where c describes the effective admixture of the B1-lattice potentials
compiled in Table~\ref{Table1} as well. Keeping in mind that the B1
lattice is free from charges on $\alpha$ sites, there are two $\alpha$
sites per unit cell with the occupation probability $1\!-\!c$ on each.
All the other sites are occupied completely, with the charge
$\kappa(2c\!-\!3)$ on a Sb$^{3-}$ ion for neutrality. The changes in
$U^0_{\alpha}$ and $U^0_{\gamma}$ defined by (\ref{Aq15}) with the
potentials from Table~\ref{Table1} exhibit the principal tendency towards
the equalization of the potentials operating on structurally nonequivalent
Ni$^+$ ions, as shown in Fig.~\ref{Fig3}. It is surprising that the
potential degeneracy obtained here takes place just at the chemical
composition which is the most prominent in experiments \cite{Rand96}.

It is evident that apart from a large number of vacant states, equal
potentials on $\alpha$ and $\gamma$ sites are capable of promoting the
extraordinary high Ni diffusivity \cite{Heum66}. Basing on this potential
equality, the existence of such a state may presumably be associated with
the creation of metallic bonds in the Ni subsystem that is much more
effective if all three interpenetrating fcc sublattices composed of Ni$^+$
ions contribute evenly to their formation. A large number of metallic
bonds can in turn depress the effect of local vacancies on the conduction
electrons. As a result, the charge transfer determining the actual value
of $\kappa$ is expected to be the same for all Ni$^+$ ions, in support of
the uniform description developed above.

As anticipated, the same trend is inherent in metallic alloys of relative
structure \cite{Robe91,Redd01,Mano99}. The fact that the Coulomb
contribution is to be less pronounced there may be regarded as one more
evidence in favor of the potential equalization discussed.

It is worth noting that the present consideration of the effect of
relative potentials is nonetheless based on absolute values of local
potentials in point-charge lattices \cite{Khol01,Kho02b}, so that their
combination in form (\ref{Aq15}) is informative. On the other hand,
keeping in mind that the value of $U^0_{\beta}($D$0_3)$ is, in absolute
units, close to $U^0_{\mathrm{As}}$, but the number of cations is larger,
the effect of $\mPhi_{\mathrm{top}}^{\mathrm{deloc}}$ defined like
(\ref{Aq13}) violates relation (\ref{Aq8}) for anion states and so renders
the stoichiometric Ni$_3$Sb compound unstable, in agreement with
experiments \cite{Rand96}. Dynamical effects associated with Ni
diffusivity and contributing to $\mPhi_{\mathrm{top}}^{\mathrm{deloc}}$ so
as to restore (\ref{Aq8}) at finite $c$ will be discussed elsewhere.

\begin{acknowledgments}
I am grateful to Professor V.~L. Ginzburg, Professor E.~G. Maksimov and
Professor V.~G. Vaks for their encouragement.
\end{acknowledgments}

\end{document}